\documentclass{elsart41}
\usepackage{graphics}
\usepackage{graphicx}
\usepackage{amssymb}
\begin{document}

\begin{frontmatter}



\title{Thermodynamic Properties of (Ca,Sr)$_2$RuO$_4$ in Magnetic Fields}


\author[cgn]{J.~Baier\corauthref{baier}},
\ead{baier@ph2.uni-koeln.de}
\author[cgn]{T.~Zabel},
\author[cgn]{M.~Kriener},
\author[cgn]{P.~Steffens},
\author[cgn]{O.~Schumann},
\author[cgn]{O.~Friedt},
\author[cgn]{A.~Freimuth},
\author[paris]{A.~Revcolevschi},
\author[kyoto]{S.~Nakatsuji},
\author[kyoto]{Y.~Maeno},
\author[cgn]{T.~Lorenz}, and
\author[cgn]{M.~Braden}

\address[cgn]{II. Physikalisches Institut, University of Cologne, 50937 Cologne, Germany}
\address[paris]{Lab. de Physico-Chimie de  l'\'Etat Solide, Universit\'e Paris-Sud, France}
\address[kyoto]{Department of Physics, Kyoto University, Kyoto 606-8502, Japan}

\corauth[baier]{Corresponding author}



\begin{abstract}
We have studied the influence of a magnetic field on the
thermodynamic properties of Ca$_{2-x}$Sr$_{x}$RuO$_4$ in the
intermediate metallic region with tilt and rotational distortions
($0.2\leq x \leq 0.5$). We find strong and anisotropic thermal
expansion anomalies at low temperatures, which are suppressed and
even reversed by a magnetic field. The metamagnetic transition of
Ca$_{1.8}$Sr$_{0.2}$RuO$_4$ is accompanied by a large
magnetostriction. Furthermore, we observe a strong magnetic-field
dependence of $c_p/T$, that can be explained by magnetic
fluctuations.
\end{abstract}

\end{frontmatter}

The phase diagram of Ca$_{2-x}$Sr$_{x}$RuO$_4$ possesses a rich
spectrum of structural distortions accompanied by changes of the
physical properties \cite{nakatsuji00a,nakatsuji00b,friedt01}. The
spin-triplet superconductor Sr$_2$RuO$_4$ is driven to the
antiferromagnetic Mott insulator Ca$_2$RuO$_4$. With decreasing Sr
content, the K$_2$NiF$_4$ structure of pure Sr$_2$RuO$_4$ gets more
and more distorted. First, a rotation of the RuO$_6$ octahedra
around the $c$ axis occurs. This rotational distortion comes along
with an increase of the low-T susceptibility $\chi$, that is about a
factor 200 larger at $x=0.5$ than that of Sr$_2$RuO$_4$
\cite{nakatsuji00a}. Below $x=0.5$ an additional tilt of the
octahedra around an in-plane axis occurs \cite{friedt01} and the
low-field magnetization at low temperatures is reduced again. In the
tilted phase, at $x=0.2$, a metamagnetic transition at $H_c\simeq
5$\,T is observed, leading to a high-field magnetization which
exceeds that of Ca$_{1.5}$Sr$_{0.5}$RuO$_4$ \cite{nakatsuji03a}. The
variation of the low-T specific heat divided by temperature $c_p/T$
as a function of Ca content resembles that of $\chi$. First, $c_p/T$
increases upon Ca-doping and reaches an unusually high value at
$x=0.5$ ($c_p/T\simeq 255$\,mJ/mole\,Ru$\cdot$K$^2$), that lies in
the range of typical heavy fermion compounds \cite{nakatsuji03a}.
Upon further increase of the Ca content, $c_p/T$ again decreases and
shows a non-monotonous temperature dependence at low temperatures for
$x=0.2$.

Here, we present measurements of the magnetostriction and the
magnetic-field dependence of thermal expansion and specific heat of
two single crystals, Ca$_{1.8}$Sr$_{0.2}$RuO$_4$ ($x=0.2$) and
Ca$_{1.5}$Sr$_{0.5}$RuO$_4$ ($x=0.5$), that were grown by a floating
zone technique at Kyoto University ($x=0.2$) and at the Universit\'e
of Paris Sud ($x=0.5$). Magnetostriction and thermal expansion were
recorded by high-resolution capacitive dilatometry and the specific
heat was measured between $300$\,mK and $60$\,K by a home-built
Nernst calorimeter based on a commercial $^3$He refrigerator. The
measurements were carried out in magnetic fields up to 14\,T.
\begin{figure}[tb]
\begin{center}
\includegraphics[clip,width=0.45\textwidth]{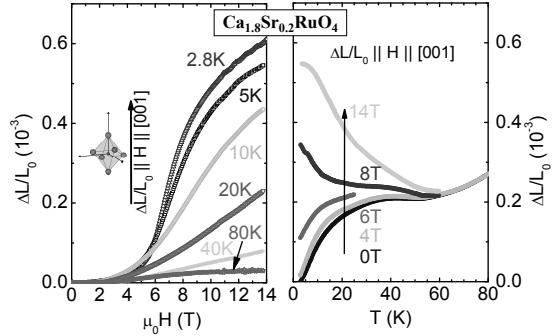}
\end{center}
\caption{Magnetostriction (left) and thermal expansion (right) of
Ca$_{1.8}$Sr$_{0.2}$RuO$_4$ ($x=0.2$). The length change was recorded
parallel to the $c$ axis in a longitudinal magnetic field.}
\label{fig1}
\end{figure}
\begin{figure}[tb]
\begin{center}
\includegraphics[clip,width=0.45\textwidth]{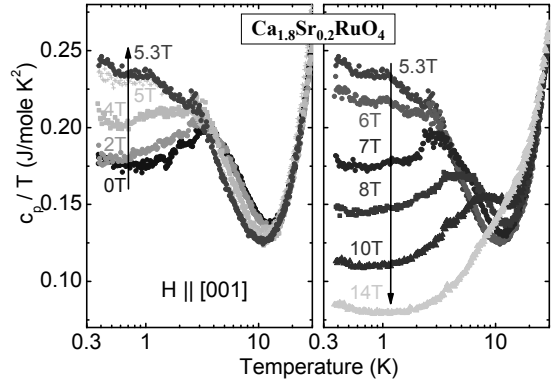}
\end{center}
\caption{Specific heat divided by temperature $c_p/T$ of
Ca$_{1.8}$Sr$_{0.2}$RuO$_4$ in magnetic fields below (left) and
above (right) the critical field of the metamagnetic transition.}
\label{fig2}
\end{figure}
Ca$_{1.8}$Sr$_{0.2}$RuO$_4$ exhibits a large magnetostriction with a
steep increase around $6$\,T at the metamagnetic transition at
lowest temperatures (see fig.\ref{fig1}, left panel). The length
change has been recorded parallel to the $c$ axis, in a longitudinal
applied field. On the same sample we find an anomalous thermal
expansion (see fig.\ref{fig1}, right panel). Upon cooling below
25\,K, a strong compression of the $c$ axis sets in. This anomalous
compression is systematically reduced by magnetic fields with $H\leq
6$\,T. Above $6$\,T, this compression turns into a $c$ axis
expansion. Measurements of the thermal expansion perpendicular to
the $c$ axis (not shown) reveal an opposed behavior, with the $ab$
plane shrinking upon cooling below 25\,K \cite{kriener04b}. A similar
thermal expansion anomaly is also observed beyond the tilted phase
for samples with $x=0.5, 0.62$ and $1$, but gets more pronounced
with decreasing Sr content \cite{kriener04b}. This low-T thermal
expansion anomaly and the effect of the magnetic field have to be
attributed to an electronic origin. A detailed discussion and the
results of neutron diffraction experiments on the same samples have
been presented in a previous study
\cite{kriener04b}.

We observe also a strong field-dependence of the low-T specific
heat. As mentioned above, without magnetic field, $c_p/T$ is reduced
for $x=0.2$ in comparison with $x=0.5$. Additionally, it possesses a
non-monotonous temperature dependence, with a maximum at about 5\,K.
As presented in fig.~\ref{fig2}, the application of a magnetic field
leads first to an increase of the low-T value of $c_p/T$. At a field
of 5.3\,T, in the vicinity of the metamagnetic transition, it reaches
a maximum value close to the $c_p/T$ value for $x=0.5$ at $H=0$\,T. A
further increase of the magnetic field leads again to a decrease of
$c_p/T$. This suggests that the enhancement of $c_p/T$ at the
metamagnetic transition is due to magnetic fluctuations that are
suppressed by the magnetic field in the spin-polarized phase at
$H>H_c$. Recently, it was shown by inelastic neutron scattering
experiments on Ca$_{1.38}$Sr$_{0.62}$RuO$_4$, that there are strong
magnetic fluctuations in this compound \cite{friedt03a}.
\begin{figure}[tb]
\begin{center}
\includegraphics[clip,width=0.45\textwidth]{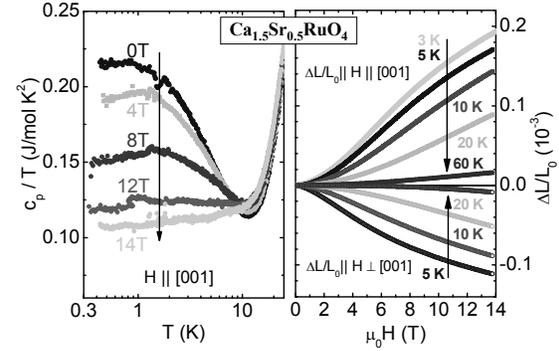}
\end{center}
\caption{Left: $c_p/T$ of Ca$_{1.5}$Sr$_{0.5}$RuO$_4$ in magnetic
fields applied parallel to the $c$ axis. Right: Magnetostriction of
Ca$_{1.5}$Sr$_{0.5}$RuO$_4$ parallel and perpendicular to the $c$
axis in a longitudinal applied field in each case.} \label{fig3}
\end{figure}

Ca$_{1.5}$Sr$_{0.5}$RuO$_4$ shows a qualitatively similar
magnetostriction, which, however, is smaller and lacks the steep
increase. The $c$ axis expands and the $ab$ plane shrinks in a
longitudinal applied magnetic field (see fig.~\ref{fig3}, right
panel). The specific heat measurements provide a monotonic decrease
of $c_p/T$ at low temperatures as a function of magnetic field (see
fig.~\ref{fig3}, left panel), similar that of
Ca$_{1.8}$Sr$_{0.2}$RuO$_4$ at $H>H_c$. These findings agree with
the idea that the increase of the Sr concentration suppresses the
metamagnetic transition field; from $\simeq 5$\,T for $x=0.2$ it goes
down close to zero for $x=0.5$.


In summary, we have found a large magnetostriction and a strong
effect of the magnetic field on the low temperature specific heat.
These data point to strong magnetic fluctuations around the
metamagnetic transition in Ca$_{2-x}$Sr$_{x}$RuO$_4$.

This work was supported by the DFG via SFB 608.

\bibliographystyle{prsty}

\end{document}